\documentclass[fleqn,usenatbib]{mnras}

\usepackage[T1]{fontenc}

\DeclareRobustCommand{\VAN}[3]{#2}
\let\VANthebibliography\thebibliography
\def\thebibliography{\DeclareRobustCommand{\VAN}[3]{##3}\VANthebibliography}

\usepackage{graphicx}
\usepackage{amsmath}	
\usepackage{amssymb}	
\usepackage{bm}
\usepackage{hyperref}
\usepackage[normalem]{ulem}
\usepackage[skip=10pt]{caption}
\usepackage{newtxtext}
\usepackage[varvw]{newtxmath} 

\let\vec\bm
\newcommand{\uvec}[1]{\vec{\hat{#1}}}

\newcommand{\diff}{\ensuremath{\mathrm{d}}}
\newcommand{\e}{\mathrm{e}}
\newcommand{\I}{\mathrm{i}}

\newcommand{\ddelta}{\delta_D}
\newcommand{\ddeltan}{\delta_D^3}

\def\vx{\vec{x}}
\def\vv{\vec{v}}
\def\vu{\vec{u}}
\def\vk{\vec{k}}
\def\va{\vec{a}}

\def\uvk{\uvec{k}}

\title[Analytic gravitational perturbations]{An analytic description of substructure-induced\\ gravitational perturbations in stellar systems}

\author[M. S. Delos]{
M. Sten Delos\thanks{E-mail: mdelos@carnegiescience.edu}
\\
The Observatories of the Carnegie Institution for Science, 813 Santa Barbara Street, Pasadena, CA 91101, USA
}

\date{Accepted XXX. Received YYY; in original form ZZZ}

\pubyear{2023}

\begin{document}
\label{firstpage}
\pagerange{\pageref{firstpage}--\pageref{lastpage}}
\maketitle

\begin{abstract}
Perturbations to stellar systems can reflect the gravitational influence of dark matter substructures. On scales much smaller than the size of a stellar system, we point out analytic connections between the stellar and dark matter distributions. In particular, the density and velocity power spectra of the stars are proportional to the density power spectrum of the perturbing dark matter, scaled by $k^{-4}$. This relationship allows easy evaluation of the suitability of a stellar system for detecting dark substructure. As examples, we show that the Galactic stellar halo is expected to be sensitive to cold dark matter substructure at wavenumbers $k\lesssim 0.3$ kpc$^{-1}$, and the Galactic disk might be sensitive to substructure at wavenumbers $k\sim 4$ kpc$^{-1}$. The perturbations considered in this work are short-lived, being rapidly erased by the stellar velocity dispersion, so it may be possible to attribute a detection to dark matter substructure without ambiguity.
\end{abstract}




\section{Introduction}

Dark matter has no known nongravitational interaction with Standard Model particles. Consequently, an important experimental strategy is to infer the properties of the dark matter from its gravitational influence alone. If the dark matter is cold, it should cluster on scales much smaller than galaxies, forming collapsed and tightly bound haloes that contain no visible matter at all. Such dark haloes could be detected gravitationally, but unambiguous detection remains elusive.

A longstanding approach is to search for the gravitational influence of dark structures on visible systems of stars.
Heating of dwarf galaxies by dark haloes has been considered \citep{2015A&A...575A..59S}, as has heating of star clusters \citep{2019MNRAS.484.5409P,2019MNRAS.488.5748W}, of tidal streams of stars \citep{2002MNRAS.332..915I,2002ApJ...570..656J,2009ApJ...705L.223C,2023arXiv230108991C}, and of binary star systems \citep{2010arXiv1005.5388P,2013JCAP...03..001G,2023MNRAS.525.5813R}.
We focus on a refinement of this approach, which is to consider inhomogeneity induced by the gravitational influence of the dark structures.
Tidal streams are the most common target for this strategy \citep{2008ApJ...681...40S,2011ApJ...731...58Y,2012ApJ...748...20C,2014ApJ...788..181N,2015MNRAS.454.3542E,2016ApJ...820...45C,2016MNRAS.463..102E,2016MNRAS.457.3817S,2017MNRAS.466..628B,2018JCAP...07..061B,2019MNRAS.484.2009B,2019ApJ...880...38B,2020MNRAS.494.5315D,2021MNRAS.502.2364B,2021JCAP...10..043B,2021A&A...649A..55K,2021ApJ...911..149L,2021MNRAS.501..179M,2022MNRAS.513.3682D,2022ApJ...941..129D,2022AJ....163...18F,2022PDU....3500978M,2022ApJ...925..118T}, although galactic disks \citep[e.g.][]{2018MNRAS.480.4244C,2023MNRAS.521..114T} and spheroids \citep{2022A&C....4100667B,2023MNRAS.519..530D} have been considered as well.

We use analytic methods to derive how the density and velocity fields of a gravitationally perturbed stellar system, and the statistics thereof, are related to the perturbing matter distribution.
We consider arbitrary systems but focus on scales much smaller than the size of the system.
On these length scales, the velocity dispersion of the stars erases perturbations on time scales much shorter than an orbital period.
Although their short lifetimes limit the amplitudes of these perturbations, their transience can also be an advantage.
Perturbations have been detected in stellar streams, but due to their long lifetimes, it is not clear that these perturbations can be attributed to dark matter substructure \citep[e.g.][]{2005AJ....129.1906C,2008MNRAS.387.1248K,2010MNRAS.401..105K,2012MNRAS.420.2700K,2016MNRAS.463L..17A,2017NatAs...1..633P,2020ApJ...891..161I,2022MNRAS.511.2339Q,2023arXiv231001485W}. Sources of short-lived inhomogeneity are easier to disambiguate.

We find simple analytic relationships between the gravitationally induced perturbations of a stellar system and the properties of the perturbing structures. The power spectra of the stellar density and velocity fields become algebraically related to the power spectrum of the perturber density field.
By comparing these induced power spectra to the intrinsic Poisson noise associated with the number of stars, one can easily test the sensitivity of a stellar system as a probe of dark substructure.

This article is organized as follows. In section~\ref{sec:pert}, we explore how perturbations to a stellar distribution are related to the density field of the perturbing matter. In section~\ref{sec:power}, we specialize to two-point statistics, deriving relationships between the stellar density and velocity power spectra and the nonlinear matter power spectrum. In section~\ref{sec:disc}, we discuss the regimes in which these results are valid, and we explore some applications. We conclude in section~\ref{sec:conc}.

\section{Perturbations to the stellar distribution}\label{sec:pert}

We begin by exploring how a stellar distribution is perturbed by the gravity of a separate distribution of matter. Let $f(\vx,\vv, t)$ be the distribution function of the stars, which depends on position $\vx$, velocity $\vv$, and time $t$.
The collisionless Boltzmann equation reads
\begin{equation}\label{BE}
	\frac{\partial f}{\partial t} + \vv\cdot \frac{\partial f}{\partial \vx} + \va(\vx,t)\cdot\frac{\partial f}{\partial \vv}=0,
\end{equation}
where $\va(\vx,t)$ is the gravitational acceleration at the position $\vx$ and time $t$.
Now separate $f(\vx,\vv,t)=f_0(\vv)+f_1(\vx,\vv,t)$ into the unperturbed distribution function $f_0$, which we approximate to be spatially uniform and constant in time, and the perturbation $f_1$. At linear order in the perturbations, we obtain
\begin{equation}\label{f1eqx}
	\frac{\partial f_1(\vx,\vv,t)}{\partial t}
	+ \vv\cdot\frac{\partial f_1(\vx,\vv,t)}{\partial \vx}
	+ \va(\vx,t)\cdot \frac{\partial f_0(\vv)}{\partial \vv}
	=0.
\end{equation}
Note that we approximate the acceleration $\va(\vx,t)$ to be entirely perturbative.

To solve this, we Fourier transform from position $\vx$ to wavenumber $\vk$, defining $f_1(\vk,\vv,t)\equiv \int\diff^3\vec x \,\e^{-\I\vk\cdot\vx}f_1(\vx,\vv,t)$ and $\va(\vk,t)$ similarly. Equation~(\ref{f1eqx}) becomes
\begin{equation}\label{f1eq}
	\frac{\partial f_1(\vk,\vv,t)}{\partial t}
	+ \I \vk\cdot\vv f_1(\vk,\vv,t)
	+ \va(\vk,t)\cdot \frac{\partial f_0(\vv)}{\partial \vv}
	=0,
\end{equation}
the solution to which is
\begin{align}\label{DF}
	f_1(\vk,\vv,t) = 
	-\int_{t_0}^t\diff t^\prime 
	\e^{-\I\vk\cdot\vv(t-t^\prime)}
	\va(\vk,t^\prime)\cdot
	\frac{\partial f_0(\vv)}{\partial \vv}
\end{align}
for an arbitrary integration constant $t_0$, which can be interpreted as the time that perturbations began.
We will take $t_0\to-\infty$, i.e., we assume that perturbations began in the arbitrarily distant past.

We now approximate that the unperturbed stellar velocity distribution is Maxwellian with dispersion $\sigma_*$ per dimension, so that
\begin{align}\label{DF0}
	f_0(\vv) = \frac{\bar n_*}{(2\pi\sigma_*^2)^{3/2}}
	\exp\!\left(-\frac{\vv^2}{2\sigma_*^2}\right).
\end{align}
Here, $n_*\equiv\int \diff^3\vv f_0(\vv)$ is the unperturbed stellar number density.
We also specialize to the gravitational acceleration
\begin{equation}
	\vec a(\vec k,t)= 4\pi\I G(\vk/k^2)\rho(\vk,t)
\end{equation}
sourced by a mass density field $\rho(\vec k,t)$.
Then the perturbation to the distribution function, given by equation~(\ref{DF}), becomes
\begin{align}\label{DF1}
	f_1(\vk,\vv,t)
	&=4\pi\I G
	\frac{\vec k\cdot\vv}{k^2\sigma_*^2}
	f_0(\vv)
	\int_{-\infty}^t\diff t^\prime 
	\e^{-\I\vk\cdot\vv(t-t^\prime)}
	\rho(\vk,t^\prime).
\end{align}

\subsection{Density perturbations}

The fractional stellar density perturbation is
\begin{align}
	\delta_*(\vk,t)
	&=
	\frac{1}{\bar n_*}\int\diff^3\vv\, f_1(\vk,\vv,t)
	\nonumber\\\label{deltastar}
	&=
	4\pi G
	\int_{-\infty}^t\diff t^\prime
	(t-t^\prime)\e^{-\frac{1}{2}k^2\sigma_*^2(t-t^\prime)^2}
	\rho(\vec k,t^\prime).
\end{align}
The interpretation of this equation is simple if we imagine how the stars would respond to an external density field that is only present at one moment $t=0$, so that  $\rho(\vk,t^\prime)\propto\ddelta(t^\prime)$, where $\ddelta$ is the Dirac delta function. The resulting density perturbation is spatially in phase with $\rho$ but evolves in time as $\delta_*\propto t\,\e^{-(k\sigma_* t)^2/2}$. This means it grows initially linearly until $t\sim(k\sigma_*)^{-1}$, at which point it is exponentially suppressed by the stellar velocity dispersion. The integral in equation~(\ref{deltastar}) simply sums the outcomes from a continuum of such instantaneous gravitational sources.

It can be convenient to eliminate the integral over the matter density field $\rho$ by temporal Fourier transformation from time $t$ to frequency $\omega$. In wavenumber-frequency space, the density perturbation $\delta_*(\vk,\omega)\equiv\int_{-\infty}^\infty\diff t\,\e^{\I\omega t}\delta_*(\vk,t)$ becomes
\begin{align}\label{deltafreq0}
	\delta_*(\vk,\omega)
	&=
	4\pi G
	\int_{-\infty}^\infty\frac{\diff\omega^\prime}{2\pi}
	\rho(\vec k,\omega^\prime)
	\nonumber\\&\hphantom{=}\times
	\int_{-\infty}^\infty\diff t
	\int_{-\infty}^t\diff t^\prime
	(t-t^\prime)\e^{\I\omega t-\I\omega^\prime t^\prime-\frac{1}{2}k^2\sigma_*^2(t-t^\prime)^2}
\end{align}
in terms of the similarly transformed matter density field $\rho(\vec k,\omega)$.
Substituting $\Delta t\equiv t-t^\prime$ separates the time integrals into a product of
$\int_{-\infty}^\infty\diff t^\prime\,\e^{\I(\omega-\omega^\prime)t^\prime}=2\pi\ddelta(\omega-\omega^\prime)$ and $\int_0^\infty\diff\Delta t\,\Delta t\,\e^{\I\omega\Delta t-\frac{1}{2}k^2\sigma_*^2\Delta t^2}$. With some manipulation, we obtain
\begin{align}\label{deltafreq1}
	\delta_*(\vk,\omega)
	&=
	\frac{4\pi G}{k^2\sigma_*^2} g\!\left(\frac{\omega}{k\sigma_*}\right)\rho(\vec k,\omega),
\end{align}
where the frequency dependence is contained in the function
\begin{align}\label{gdef}
	g(x)
	&\equiv \int_0^\infty\diff y \, y \, \e^{-\frac{1}{2}y^2+\I x y}.
\end{align}
For numerical evaluation, it is useful to note that
\begin{align}\label{g}
	g(x)
	&=1 - \sqrt{2}\,x\, F(x/\sqrt{2}) + \I\sqrt{\pi/2}\,x\,\e^{-x^2/2},
\end{align}
where $F(x)\equiv \e^{-x^2}\int_0^x\diff y\,\e^{y^2}$ is the Dawson integral.

\begin{figure}
	\centering
	\includegraphics[width=\columnwidth]{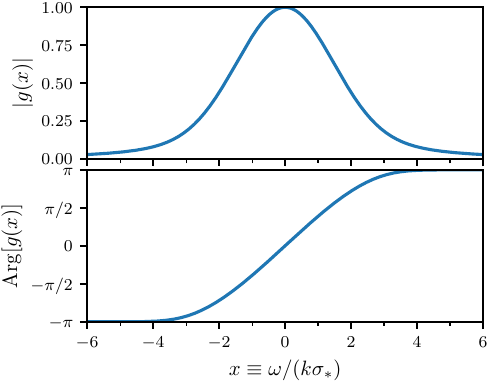}
	\caption{
		The function $g(x)$ appearing in the relationship in equation~(\ref{deltafreq1}) between the stellar density perturbation $\delta_*(\vk,\omega)$ and the matter density field $\rho(\vk,\omega)$. Here $x\equiv\omega/(k\sigma_*)$ and $\sigma_*$ is the stellar velocity dispersion per dimension.
		The upper panel shows the magnitude of $g(x)$, while the lower panel shows its complex argument.
		For low frequencies $|\omega|\ll k\sigma_*$, $g(x)\simeq 1$, so the stellar and matter fields are in phase. 
		For high frequencies $|\omega|\gg k\sigma_*$, $g(x)\simeq -1/x^2$, so the stellar and matter fields are maximally out of phase.
	}
	\label{fig:transfer}
\end{figure}

We plot $g(x)$ in figure~\ref{fig:transfer}. Note that $g(x)$ ranges from $1$ for $|x|\ll 1$ to $-1/x^2$ for $|x|\gg 1$. Thus, in the low-frequency limit,
\begin{align}\label{deltafreqlow}
	\delta_*(\vk,\omega)
	&\simeq
	\frac{4\pi G}{k^2\sigma_*^2}\rho(\vec k,\omega), & \text{ if \ \ } |\omega|\ll k\sigma_*.
\end{align}
Recall that $(k\sigma_*)^{-1}$ is the time scale over which the velocity dispersion erases stellar perturbations on the scale $k^{-1}$. The interpretation of equation~(\ref{deltafreqlow}) is that if the driving density field $\rho$ changes only on much longer time scales $|\omega|^{-1}\gg (k\sigma_*)^{-1}$, then the stellar response $\delta_*$ aligns with the density field in space and time but is suppressed by the factor $(k\sigma_*)^{-2}$.
Conversely, in the high-frequency limit,
\begin{align}\label{deltafreqhigh}
	\delta_*(\vk,\omega)
	&\simeq
	-\frac{4\pi G}{\omega^2}\rho(\vec k,\omega), & \text{ if \ \ } |\omega|\gg k\sigma_*.
\end{align}
This means that if the driving density field $\rho$ changes on much shorter time scales $|\omega|^{-1}\ll (k\sigma_*)^{-1}$, then the stellar response $\delta_*$ is opposite from $\rho$ in phase and is suppressed by the factor $|\omega|^{-2}$.

\subsection{Velocity perturbations}

Through similar calculations, or by using the continuity equation, the mean velocity of the stars is
\begin{align}
	\bar \vv(\vk,t)
	\label{vstar}
	&=
	4\pi\I G\frac{\vk}{k^2}
	\!\!\int_{-\infty}^t\!\!\!\!\!\diff t^\prime
	\left[1-k^2 \sigma_*^2(t\!-\!t^\prime)^2\right]
	\e^{-\frac{1}{2}k^2\sigma_*^2(t-t^\prime)^2}\!\!
	\rho(\vk,t^\prime)
\end{align}
in wavenumber space and
\begin{align}\label{vfreq1}
	\bar\vv(\vk,\omega)
	&=
	\frac{4\pi G\omega \vk}{k^4\sigma_*^2} g\!\left(\frac{\omega}{k\sigma_*}\right)\rho(\vec k,\omega)
\end{align}
in wavenumber-frequency space.

If the stellar distribution $\delta_*(\vk,\omega)$ were known, then the mass distribution $\rho(\vk,\omega)$ that is gravitationally driving it could be inferred using equation~(\ref{deltafreq1}), and there would be no need to consider $\bar \vv$. However, knowledge of $\delta_*(\vk,\omega)$ means knowing the stellar distribution at all times. In practice, we can measure stellar positions and velocities today only. If we define the present day to be $t=0$, then
\begin{align}\label{deltak1}
	\delta_*(\vk)
	&=
	\frac{4\pi G}{k^2\sigma_*^2} 
	\int_{-\infty}^\infty\frac{\diff\omega}{2\pi}
	g\!\left(\frac{\omega}{k\sigma_*}\right)\rho(\vec k,\omega),
	\\\label{vk1}
	\bar\vv(\vk)
	&=
	\frac{4\pi G \vk}{k^4\sigma_*^2} 
	\int_{-\infty}^\infty\frac{\diff\omega}{2\pi}
	g\!\left(\frac{\omega}{k\sigma_*}\right)\omega\rho(\vec k,\omega)
\end{align}
at the present time. That is, by measuring the present-day density and velocity fields of the stellar distribution, it is possible in principle to observationally infer the zeroth and first frequency moments of the quantity $g[\omega/(k\sigma_*)]\rho(\vec k,\omega)$.

\section{Stellar and dark matter power spectra}\label{sec:power}

Further exploration of the possibility of probing the full matter density field $\rho(\vk,\omega)$ lies beyond the scope of this article. Instead, we will focus on the simpler possibility of probing its power spectrum.
Suppose that the gravitating matter distribution $\rho$ is statistically homogeneous (in space) and constant (in time). We can define a ``spatiotemporal'' matter density power spectrum $\mathcal{P}_\rho(\vk,\omega)$ such that
\begin{equation}\label{PP}
	(2\pi)^4\ddeltan(\vk-\vk^\prime)\ddelta(\omega-\omega^\prime)\mathcal{P}_\rho(\vk,\omega)\equiv\langle\rho(\vk,\omega)\rho^*(\vk^\prime,\omega^\prime)\rangle.
\end{equation}
Here $\rho^*$ is the complex conjugate of $\rho$, the angle brackets denote the average, and $\ddelta$ and $\ddeltan$ are the one- and three-dimensional Dirac delta functions, respectively.
Predictions for $\mathcal{P}_\rho(\vk,\omega)$ can be estimated directly from numerical simulations or semianalytic substructure models, which is a subject of ongoing work. Here we approximate that the stellar perturbations arise entirely from dark matter, i.e. that $\rho$ is the dark matter density field, and we discuss how $\mathcal{P}_\rho(\vk,\omega)$ connects to conventional descriptions of dark matter structure.

As we will see, it is useful to define a function $Q_\rho(\vk,\omega)$ such that
\begin{align}\label{Qdef}
	\mathcal{P}_\rho(\vk,\omega)
	&\equiv
	Q_\rho(\vk,\omega)P_\rho(k).
\end{align}
Here $P_\rho(k)$ is the spatial matter density power spectrum defined such that
\begin{equation}\label{P}
	\langle \rho(\vk)\rho^*(\vk^\prime)\rangle \equiv (2\pi)^3 \ddeltan(\vk-\vk^\prime)P_\rho(k),
\end{equation}
where $\rho(\vk)$ is the spatial Fourier transform of the density field at some fixed time, which we can take to be $t=0$. $P_\rho(k)$ is constant in time by assumption. Also, we assume statistical isotropy at fixed times, so $P_\rho(k)$ does not depend on the direction of $\vk$.

Note that $P_\rho(k)=\bar\rho^2 P(k)$, where $P(k)$ is the power spectrum of the density contrast with respect to the cosmological average $\bar\rho$, a commonly discussed quantity in cosmology.
However, we emphasize that $P_\rho(k)$ is not the matter power spectrum extrapolated using linear-order cosmological perturbation theory (as is standard in cosmology) but rather the nonlinear matter power spectrum, which describes the nonlinearly evolved matter density field.

\subsection{Dark matter substructure}

Nonlinear dark matter structure is often discussed in terms of subhalo models, and $P_\rho(k)$ can be evaluated from such models in the following way. Suppose that haloes of mass $M$ have differential number density $\diff n/\diff M$ per mass interval and spatial volume and that a halo of mass $M$ is spherical with density profile $\rho(r|M)$ as a function of radius $r$.
Then
\begin{equation}\label{Phalo}
	P_\rho(k)=\int\diff M\frac{\diff n}{\diff M} \left|\rho(k|M)\right|^2
\end{equation}
if halo positions are uncorrelated, where $\rho(k|M)$ is the three-dimensional Fourier transform of $\rho(r|M)$, i.e., 
\begin{equation}\label{profile}
	\rho(k|M) = \int_0^\infty 4\pi r^2\diff r\, \rho(r|M)\frac{\sin(kr)}{kr}.
\end{equation}
If haloes have a nontrivial spatial distribution, then equation~(\ref{Phalo}) also includes a two-halo term \citep[e.g.][]{1991ApJ...381..349S}. However, subhalo positions are expected to be mostly uncorrelated due to phase mixing and the subdominance of their mutual gravitation.

Returning to the spatiotemporal matter power spectrum $\mathcal{P}_\rho(\vk,\omega)$, the function $Q_\rho=\mathcal{P}_\rho/P_\rho$ (equation~\ref{Qdef}) can be estimated as follows.
Suppose the dark matter field were moving coherently at a velocity $\vec u$.
Then as a function of time, $\rho(\vk,t)=\rho(\vk)\e^{-\I \vec k\cdot\vec u t}$. Fourier transforming in time leads to
\begin{equation}
	\rho(\vec k,\omega)=2\pi\ddelta(\vec k\cdot\vec u-\omega)\rho(\vec k),
\end{equation}
and so
\begin{align}
	\langle\rho(\vec k,\omega)\rho^*(\vec k^\prime,\omega^\prime)\rangle
	&=
	(2\pi)^5\ddelta(\vec k\cdot\vec u-\omega)\ddelta(\vec k^\prime\cdot\vec u-\omega^\prime)
	\nonumber\\&\hphantom{=}\times
	\ddeltan(\vec k-\vec k^\prime)P_\rho(k).
\end{align}
For the simple scenario of coherent motion, evidently
\begin{align}\label{Q_u}
	Q_\rho(\vec k,\omega)=2\pi\ddelta(\vec k\cdot\vec u-\omega).
\end{align}
Notice how the spatiotemporal power spectrum $\mathcal{P}_\rho=Q_\rho P_\rho$ can be anisotropic even if the constant-time spatial power spectrum $P_\rho$ is isotropic.

A more realistic case is where the dark matter substructure has a Maxwellian velocity distribution with dispersion $\sigma$ per dimension.
If density fields moving in different directions are uncorrelated, it suffices to average equation~(\ref{Q_u}) over this velocity distribution, yielding
\begin{align}
	Q_\rho(\vec k,\omega)
	&=
	\int\diff^3\vu \frac{\e^{-u^2/(2\sigma^2)}}{(2\pi\sigma^2)^{3/2}}2\pi\ddelta(\vk\cdot\vu-\omega)
	\nonumber\\\label{Q_s}&=
	\frac{\sqrt{2\pi}}{k\sigma}\e^{-\frac{\omega^2}{2k^2\sigma^2}}
\end{align}
\citep[see also the more careful discussion in][]{2022MNRAS.513.3682D}.
This description is appropriate for studying perturbations in a stellar distribution that is at rest with respect to the dark matter distribution, such as the stars in the spheroidal component of a galaxy.

A further sophistication is to suppose that the stellar distribution moves at velocity $\vu_0$ with respect to the centre of the Maxwellian dark matter velocity distribution. Now if $\vu$ are the velocities in the rest frame of the stellar distribution, then $\vu^\prime\equiv\vu+\vu_0$ are Maxwellian distributed. In this case,
\begin{align}
	Q_\rho(\vec k,\omega)
	\label{Q_su}&=
	\frac{\sqrt{2\pi}}{k\sigma}\e^{-(\omega+\vk\cdot\vu_0)^2/(2k^2\sigma^2)}.
\end{align}
This description is appropriate when studying perturbations to stars in a galactic disk or in a satellite galaxy, for example, since these systems exhibit net motion with respect to the dark matter substructure.

\subsection{Stellar density power}\label{sec:Pstar}

Consider now the spatial power spectrum $P_*(\vk)$ of the stellar density contrast, defined similarly to equation~(\ref{P}) but potentially anisotropic. From equations (\ref{deltak1}), (\ref{PP}), and~(\ref{Qdef}), $P_*(\vk)$ is related to the dark matter density power spectrum $P_\rho(k)$ as
\begin{align}\label{Pstar0}
	P_*(\vk)
	&=
	\frac{(4\pi G)^2}{k^3\sigma_*^3} 
	\left[
	\int_{-\infty}^\infty\frac{\diff x}{2\pi}
	\left|g(x)\right|^2 Q_\rho(\vk,x k\sigma_*)
	\right]
	P_\rho(k)
\end{align}
with $g(x)$ again given by equation~(\ref{g}). In general, the integral must be evaluated numerically. Note that $|g(x)|$ is numerically well behaved (see figure~\ref{fig:transfer}), as is $Q_\rho$ in realistic cases (equations \ref{Q_s} and~\ref{Q_su}).
However, there are some cases where the relationship between $P_*$ and $P_\rho$ can be written analytically.
For this purpose, it is useful to note that equation~(\ref{gdef}) implies
\begin{align}\label{absg2}
	|g(x)|^2 &= \int_0^\infty \diff y\int_0^\infty \diff z \, y\, z\, \e^{-\frac{1}{2}(y^2+z^2)+\I x (y-z)}.
\end{align}

\subsubsection{Maxwellian dark matter substructure}\label{sec:Pstar_s}

\begin{figure}
	\centering
	\includegraphics[width=\columnwidth]{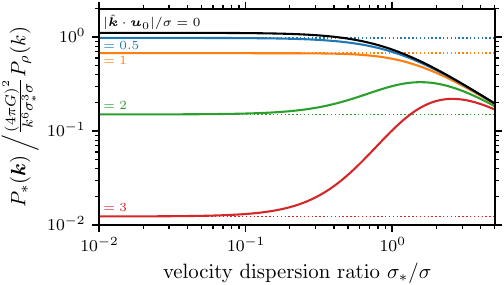}
	\caption{Integral factor between the stellar density power spectrum $P_*$ and the matter density power spectrum $P_\rho$. This is the bracketed factor in equation~(\ref{Pstar2}), and it depends on the ratio $\sigma_*/\sigma$ between the stellar and matter velocity dispersions (horizontal axis) and on $\uvk\cdot\vu_0/\sigma$ (different colours), the ratio between the projection of the mean stellar velocity onto the wavevector and the matter velocity dispersion. The $\uvk\cdot\vu_0=0$ case (black curve) corresponds to $I_*(s)$ given by equation~(\ref{Istar}). The dotted lines show the approximation of equation~(\ref{Pstar3}), which is valid in the limit of a cold stellar distribution.}
	\label{fig:Istar}
\end{figure}

Suppose that the dark matter substructure has a Maxwellian velocity distribution in the rest frame of the stellar distribution with dispersion $\sigma$ per dimension, so $Q_\rho$ is given by equation~(\ref{Q_s}).
Substituting equation~(\ref{absg2}) into equation~(\ref{Pstar0}) and carrying out the $x$ integral first yields
\begin{align}\label{Pstar1}
	P_*(k)
	&=
	\frac{(4\pi G)^2}{k^4\sigma_*^3\sigma} 
	I_*\!\left(\frac{\sigma_*}{\sigma}\right)
	P_\rho(k),
\end{align}
where we define the integral
\begin{align}
	I_*(s)
	&\equiv
	\frac{1}{s}
	\int_0^\infty \diff y\int_0^\infty \diff z \, y\, z\, \e^{-\frac{1}{2}(y^2+z^2)-\frac{1}{2s^2}(y-z)^2}
	\nonumber\\\label{Istar}&=
	\frac{\pi-\arctan(s\sqrt{2+s^2})}{(2+s^2)^{3/2}}+\frac{s}{2+s^2}.
\end{align}
We note that if the dark matter and the stars have the same velocity dispersion, so $\sigma_*=\sigma$, then this quantity evaluates to $I_*(1)=(9+2\pi\sqrt{3})/27\simeq 0.74$. If on the other hand the stars are much colder than the dark matter, so $\sigma_*\ll\sigma$, then it evaluates to $I_*(0)=\pi/2^{3/2}\simeq 1.11$. We plot $I_*(s)$ as the black curve in figure~\ref{fig:Istar}.

\subsubsection{Shifted Maxwellian dark matter substructure}\label{sec:Pstar_su}

Now consider dark matter substructure with the same Maxwellian velocity distribution, but suppose that the stars move at net velocity $\vu_0$ with respect to the dark matter, so that $Q_\rho$ is given by equation~(\ref{Q_su}). Then equation~(\ref{Pstar0}) becomes
\begin{align}\label{Pstar2}
	P_*(\vk)
	&=
	\frac{(4\pi G)^2}{k^4\sigma_*^3\sigma} 
	\left[
	\int_{-\infty}^\infty\!\!\frac{\diff x}{\sqrt{2\pi}}
	|g(x)|^2
	\e^{-\frac{\!(\sigma_*/\sigma)^2\!}{2}(x+\uvk\cdot\vu_0/\sigma_*)^2}
	\right]
	P_\rho(k),
\end{align}
where $\uvk\equiv\vk/k$ is the unit vector along $\vk$.
We are not able to evaluate this expression analytically, but we show the numerically evaluated integral in brackets as the coloured solid curves in figure~\ref{fig:Istar}. Note that by inspection, it can only depend on the magnitude of $\uvk\cdot\vu_0$, since $|g(x)|^2$ is an even function of $x$.

Nevertheless, we can analytically treat the following limiting scenarios. If the wavevector $\vk$ is perpendicular to the mean stellar velocity $\vu_0$, then equation~(\ref{Pstar2}) simply reduces to equation~(\ref{Pstar1}). Conversely, suppose that $|\uvk\cdot\vu_0|\gg\sigma_*$, corresponding to a cold stellar distribution with a large mean velocity that is not close to perpendicular with the wavevector $\vk$. Then $(x+\uvk\cdot\vu_0/\sigma_*)^2\simeq (\uvk\cdot\vu_0/\sigma_*)^2$, since $|g(x)|^2$ only has significant support for $|x|\sim \mathcal{O}(1)$. In this limit
\begin{align}\label{Pstar3}
	P_*(\vk)
	&\simeq
	\frac{(4\pi G)^2}{k^4\sigma_*^3\sigma}
	I_*(0)
	\e^{-\frac{1}{2}(\uvk\cdot\vu_0/\sigma)^2}
	P_\rho(k),
\end{align}
where we have used that $\int_{-\infty}^\infty\diff x |g(x)|^2=\pi^{3/2}/2=\sqrt{2\pi}I_*(0)$. As we noted above, $I_*(0)=\pi/2^{3/2}\simeq 1.11$.
Evidently, the stellar perturbations are exponentially suppressed by a nonzero mean velocity $\vu_0$. This approximation is shown in figure~\ref{fig:Istar} with the dotted curves.

\subsection{Stellar velocity power}\label{sec:Pv}

Let us define the fixed-time stellar velocity power spectrum $P_{v_i v_j}(\vk)$ such that
\begin{align}
	\langle \bar v_i(\vk)\bar v_j^*(\vk^\prime)\rangle\equiv(2\pi)^3\ddeltan(\vk-\vk^\prime)P_{v_i v_j}(\vk),
\end{align}
where $i$ and $j$ are the indices of the mean velocity vectors.
From equations (\ref{vk1}-\ref{Qdef}), we can obtain
\begin{align}\label{Pv0}
	P_{v_i v_j}(\vk)
	&=
	\frac{(4\pi G)^2 k_i k_j}{k^5\sigma_*} 
	\left[
	\int_{-\infty}^\infty\frac{\diff x}{2\pi}x^2
	|g(x)|^2 Q_\rho(\vk,x k\sigma_*)
	\right]
	P_\rho(k).
\end{align}

\subsubsection{Maxwellian dark matter substructure}\label{sec:Pv_s}

Similarly to section~\ref{sec:Pstar_s}, we can consider a Maxwellian dark matter velocity distribution in the stellar rest frame and obtain in this case
\begin{align}\label{Pv1}
	P_{v_i v_j}(\vk)
	&=
	\frac{(4\pi G)^2 k_i k_j}{k^6\sigma_*\sigma} 
	I_v\left(\frac{\sigma_*}{\sigma}\right)
	P_\rho(k),
\end{align}
where we define the function
\begin{align}\label{Iv}
	I_v(s)&\equiv
	\frac{3[\pi-\arctan(s\sqrt{2+s^2})]}{(2+s^2)^{5/2}}
	-\frac{s(1-s^2)}{(1+s^2)(2+s^2)^2}.
\end{align}
For the cases of hot ($\sigma_*=\sigma$) and cold ($\sigma_*\ll\sigma$) stellar distributions, $I_v(1)=2\pi/3^{5/2}\simeq 0.40$ and $I_v(0)=3\pi/2^{5/2}\simeq 1.67$, respectively. We plot $I_v(s)$ as the black curve in figure~\ref{fig:Iv}.

\subsubsection{Shifted Maxwellian dark matter substructure}\label{sec:Pv_su}

\begin{figure}
	\centering
	\includegraphics[width=\columnwidth]{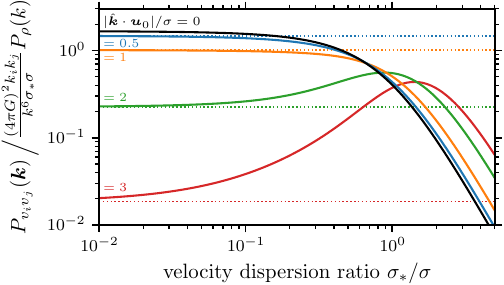}
	\caption{Integral factor between the stellar velocity power spectrum $P_{v_i v_j}$ and the matter density power spectrum $P_\rho$. This is the bracketed factor in equation~(\ref{Pv2}). As with figure~\ref{Istar}, we show how it depends on the stellar and matter velocity dispersions $\sigma_*$ and $\sigma$, respectively, and the component $\uvk\cdot\vu_0$ of the mean stellar velocity along the wavevector. The $\uvk\cdot\vu_0=0$ case (black curve) corresponds to $I_v(s)$ given by equation~(\ref{Iv}). The dotted lines show the approximation of equation~(\ref{Pv3}), which is valid in the limit of a cold stellar distribution.}
	\label{fig:Iv}
\end{figure}

Similarly to section~\ref{sec:Pstar_su}, we also consider the case where the dark matter substructure has a Maxwellian velocity distribution but the stars have a net velocity $\vu_0$ with respect to the dark matter. In this case
\begin{align}\label{Pv2}
	P_{v_i v_j}(\vk)
	&=
	\frac{(4\pi G)^2 k_i k_{\!j}\!}{k^6\sigma_*\sigma}
	\nonumber\\&\hphantom{=}\times
	\left[\!
	\int_{-\infty}^\infty\!\!\frac{\diff x}{\sqrt{2\pi}}
	x^2|g(x)|^2
	\e^{-\frac{\!(\sigma_*\!/\sigma)^2\!\!\!}{2}(x+\uvk\cdot\vu_0/\sigma_*)^2}
	\!\right]\!
	P_\rho(k),
\end{align}
where $\uvk\equiv\vk/k$ is again the unit vector along $\vk$.
We plot the bracketed integral as the coloured solid curves in figure~\ref{fig:Iv}, and note that it again depends only on the magnitude of $\uvk\cdot\vu_0$.
In the limit of a cold stellar distribution, $\sigma_*\ll|\uvk\cdot\vu_0|$, and similarly to equation~(\ref{Pstar3}) we obtain
\begin{align}\label{Pv3}
	P_{v_i v_j}(\vk)
	&\simeq
	\frac{(4\pi G)^2 k_i k_j}{k^6\sigma_*\sigma} 
	I_v(0)
	\e^{-\frac{1}{2}(\uvk\cdot\vu_0/\sigma)^2}
	P_\rho(k).
\end{align}
This approximation corresponds to the dotted curves in figure~\ref{fig:Iv}.

\subsection{Density-velocity cross-correlations}

Finally, we consider the cross power between the stellar density contrast and the mean velocity of stars. We may define a cross power spectrum $P_{*v_i}(\vk)$ such that
\begin{align}
	\langle \delta(\vk)\bar v_i^*(\vk^\prime)\rangle\equiv(2\pi)^3\ddeltan(\vk-\vk^\prime)P_{*v_i}(\vk),
\end{align}
and by equations (\ref{deltak1}-\ref{Qdef}) we obtain
\begin{align}\label{Pstarv0}
	P_{*v_i}(\vk)
	&=
	\frac{(4\pi G)^2 k_i}{k^4\sigma_*^2}
	\left[
	\int_{-\infty}^\infty\frac{\diff x}{2\pi}
	x|g(x)|^2 Q_\rho(\vk,x k\sigma_*)
	\right]P_\rho(k).
\end{align}
Note that $|g(x)|$ is symmetric in $x$, so if $Q_\rho$ is symmetric in frequency $\omega$, then $P_{*v_i}(\vk)=0$. In particular, if the dark matter has a Maxwellian velocity distribution and the stars have no net motion, so $Q_\rho$ is given by equation~(\ref{Q_s}), then there is no cross power between $\vv$ and $\delta$.

\begin{figure}
	\centering
	\includegraphics[width=\columnwidth]{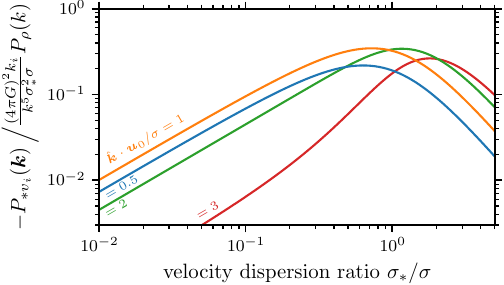}
	\caption{Integral factor between the stellar density-velocity cross power spectrum $P_{*v_i}$ and the matter density power spectrum $P_\rho$. This is the negative of the bracketed factor in equation~(\ref{Pstarv2}). It is only large when the stellar velocity dispersion $\sigma_*$, the matter velocity dispersion $\sigma$, and the component $\uvk\cdot\vu_0$ of the mean stellar velocity along the wavevector are all comparable. Note also that it is negative when $\uvk\cdot\vu_0>0$ (plotted) but positive when $\uvk\cdot\vu_0<0$ (not plotted).}
	\label{fig:Istarv}
\end{figure}

If the stars have net velocity $\vu_0$ with respect to the dark matter, then
\begin{align}\label{Pstarv2}
	P_{*v_i}(\vk)
	&=
	\frac{(4\pi G)^2 k_i}{k^5\sigma_*^2\sigma}
	\nonumber\\&\hphantom{=}\times
	\left[
	\int_{-\infty}^\infty\!\!\frac{\diff x}{\sqrt{2\pi}}
	x|g(x)|^2 \e^{-\frac{\!(\sigma_*/\sigma)^2\!}{2}(x +\uvk\cdot\vu_0/\sigma_*)^2}
	\right]\!P_\rho(k).
\end{align}
We are not able to integrate this expression analytically, nor is the $\sigma_*\ll\uvk\cdot\vu_0$ limit helpful, since it simply leads to $P_{*v_i}(\vk)=0$, but we show the numerically evaluated integral in figure~\ref{fig:Istarv}. Note that it is negative for the $\uvk\cdot\vu_0>0$ that we show, but it is odd in $\uvk\cdot\vu_0$, so it would be positive for $\uvk\cdot\vu_0<0$. However, figure~\ref{fig:Istarv} indicates that this integral is only large in magnitude when $\sigma_*\sim\sigma\sim \uvk\cdot\vu_0$, which is a physically implausible scenario. If the stars are as hot as the dark matter, then we do not also expect a comparable magnitude of coherent (e.g. rotational) motion. Thus, we generally expect $P_{*v_i}(\vk)\simeq 0$.

\section{Discussion}\label{sec:disc}

We now discuss the regimes in which the results of this analysis are applicable, and we present an application.

\subsection{Applicability}

The principal approximations made in these calculations are that the perturbing matter fields are the only source of gravitational acceleration and that the unperturbed systems are spatially uniform. Stars reside inside galaxies, of course, and we have neglected both the host galaxy's gravity and its overall density structure. These approximations are valid under the following conditions.
\begin{enumerate}
	\item The time scale $(k\sigma_*)^{-1}$ for the stellar velocity dispersion to erase perturbations is much shorter than the stars' orbital time scale. If perturbations can persist over an orbital period, then one cannot neglect orbital dynamics.
	\item The length scale $k^{-1}$ under consideration is much smaller than the overall size of the system. Otherwise, the global structure of the system must be accounted for.
\end{enumerate}
The first condition means that the stars must comprise a structure that is supported by its velocity dispersion on the scales that we are studying. The second condition does not impose an additional requirement, since the velocity dispersion sets the system's size.

Thus, the calculations in this work are applicable to spheroidal galaxies, spheroidal components of galaxies, and globular star clusters, on scales much smaller than the size of the system.
These calculations also apply to galactic disks or stellar streams, but only if $k^{-1}$ is much smaller than the system's smallest transverse size \citep[although for larger scales, one can modify the description to include an approximate account of orbital dynamics; see][]{2022MNRAS.513.3682D}.

\subsection{Sensitivity estimates}

A straightforward application of the power spectra of stellar perturbations discussed in section~\ref{sec:power} is to estimate whether gravitational perturbations would be detectable in a given stellar system.
The idea is to compare the predicted power spectrum to the intrinsic Poisson noise associated with the finite number of stars.
We focus on the stellar density power spectrum $P_*$ discussed in section~\ref{sec:Pstar}. If the stars have mean number density $\bar n_*$, then the power spectrum of the Poisson noise is $P_*=\bar n_*^{-1}$.

These estimates will be optimistic, because clustering in the stars can arise also due to the history of a stellar system. That is, stars could be clustered because they formed together or accreted together. Although we do not consider this effect further, we remark that historical clustering can be distinguished in principle from clustering associated with gravitational perturbations. When clustering arises from history, density and velocity perturbations must be strongly correlated in space; because otherwise they would rapidly dissipate. On the other hand, we showed that the transient density and velocity perturbations induced by substructure are spatially uncorrelated at a fixed time.

\subsubsection{The stellar halo at 20 kpc}

\begin{figure}
	\centering
	\includegraphics[width=\columnwidth]{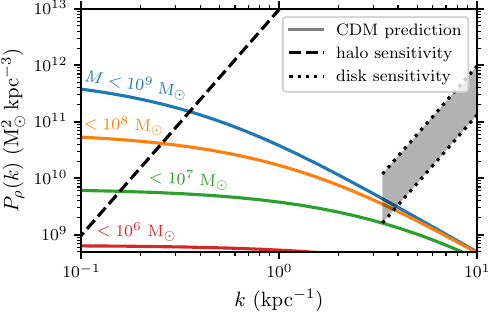}
	\caption{
		Testing the capacity of the Milky Way's stellar halo and disk to probe dark matter substructure.
		We plot the nonlinear matter power spectrum $P_\rho(k)$; the dashed and dotted curves show the sensitivity limits for the Milky Way's stellar halo at $\sim 20$~kpc and local disk stars, respectively.
		Specifically, for each wavenumber $k$, the sensitivity limit corresponds to the smallest value of $P_\rho(k)$ that would induce fluctuations in the stellar density field that exceed the level of the intrinsic Poisson noise arising from the finite number of stars.
		The sensitivity for disk stars depends on the angle (shaded band) and is strongest (lowest $P_\rho$) for wavevectors $\vk$ perpendicular to the disk rotation velocity.
		For comparison, the solid curves show predicted cold dark matter (CDM) density power spectra, with different colours corresponding to different upper limits on the subhalo mass $M$.
		We use the same subhalo model as \citet{2017MNRAS.466..628B} and \citet{2022MNRAS.513.3682D}, which is based on the CDM simulation of \citet{2008Natur.454..735D}.}
	\label{fig:halo}
\end{figure}

As a first example, we note that the Milky Way's stellar halo at a radius of about $20$~kpc has a number density of order $\bar n_*\sim 10^{-5}$~pc$^{-3}$, based on a power-law density profile with index -3.5 and local normalization of about $10^{-4}$~M$_\odot$\,pc$^{-3}$ \citep{2008A&ARv..15..145H} and assuming the average star weighs 0.5~M$_\odot$.
The power spectrum of the Poisson noise is hence $P_*(k)=\bar n_*^{-1}\sim 10^{-4}$~kpc$^{3}$.
The radial velocity dispersion is about 120~km\,s$^{-1}$ \citep{2008A&ARv..15..145H}, and if we approximate that it is isotropic and the same for stars and substructures, then equation~(\ref{Pstar1}) implies that perturber power spectra larger than $P_\rho(k)\sim 10^{13}(k/\text{kpc}^{-1})^{4}$~M$_\odot^2\,$kpc$^{-3}$ produce stellar density perturbations that exceed the level of the Poisson noise. The dashed line in figure~\ref{fig:halo} shows this sensitivity limit.

For comparison, we also show the nonlinear power spectrum that corresponds (per equations \ref{Phalo} and~\ref{profile}) to the subhalo model used by \citet{2017MNRAS.466..628B} and \citet{2022MNRAS.513.3682D} to study perturbations to stellar streams at a comparable Galactocentric radius. This subhalo model is based on the numerical simulation of \citet{2008Natur.454..735D} of a Milky Way--like dark matter halo. The subhalo mass function is taken to be $\diff n/\diff M\propto M^{-2}$ normalized so that the number density of $10^6$-$10^7$~M$_\odot$ subhaloes is $6\times 10^{-4}$~kpc$^{-3}$. Subhaloes have \citet{1990ApJ...356..359H} density profiles with scale radii $R=1.04~\text{kpc}\,(M/10^8~\mathrm{M}_\odot)^{1/2}$. This description is highly approximate, and a more accurate characterization of the local nonlinear matter power spectrum is a subject of ongoing work.

Apparently, for $k\lesssim 0.3$~kpc$^{-1}$, this substructure model would induce perturbations in the distribution of halo stars at $\sim 20$~kpc that exceed the level of the Poisson noise. We show the power spectra with different maximum subhalo masses imposed in order to test which subhaloes are driving this result. The perturbations from sub-$10^8$~M$_\odot$ subhaloes would be detectable (orange curve), but those from sub-$10^7$~M$_\odot$ subhaloes (green curve) likely are not.

\subsubsection{Local disk stars}

The dotted lines in figure~\ref{fig:halo} show the estimated sensitivity of nearby stars in the Galactic disk to perturbations by substructure, plotted only up to $k^{-1}=0.3$~kpc, which is approximately the vertical scale height of the dominant component of the disk \citep{2016ARA&A..54..529B}. We assume the stars comprise a mass density of about $0.04$~M$_\odot\,$pc$^{-3}$ and have a velocity dispersion of 30~km\,s$^{-1}$ per dimension \citep{2016ARA&A..54..529B}. We again assume that stars have average mass 0.5~M$_\odot$ and that the substructure velocity dispersion is 120~km\,s$^{-1}$ per dimension.
We adopt a rotation velocity of $u_0=250$~km\,s$^{-1}$. The sensitivity depends on the angle between the wavevector $\vk$ and the rotation velocity, which is why we show two dotted curves. Sensitivity is maximized ($P_\rho$ is lowest) when $\vk$ is perpendicular to the rotation velocity and minimized ($P_\rho$ is highest) when it is parallel.

Figure~\ref{fig:halo} suggests that dark substructure around $k\sim 4$~kpc$^{-1}$ could be barely detectable through induced density perturbations of the local stellar field. However, it should be noted that substructure that crosses the disk may be disrupted to a significantly greater degree than is reflected in the dark-matter-only simulation on which the predicted $P_\rho(k)$ in figure~\ref{fig:halo} is based \citep[e.g.][]{2010ApJ...709.1138D}.
Another concern is that the calculations in this work neglect the self-gravity of the stellar perturbations themselves, which can be relevant for galactic disks \citep[e.g.][]{2023MNRAS.522..477W}, although this effect would only amplify the stellar perturbations (and so improve the sensitivity).

\section{Conclusion}\label{sec:conc}

We derived analytic relationships between the distribution of a gravitationally perturbed stellar system and that of the matter that sources the perturbations. These results are expressed in Fourier space and are valid in the limit that the length scale $k^{-1}$ under consideration is much smaller than the size of the stellar system. We considered the full stellar density and velocity fields in section~\ref{sec:pert} and their two-point statistics in section~\ref{sec:power}.
In particular, the stellar density and velocity power spectra are related in a simple way to the nonlinear matter power spectrum. In both cases, the power spectrum of the stars is proportional to that of the perturbers, albeit weighted by $k^{-4}$. The cross power between stellar density and velocity is essentially zero.

Although the description is general, the motivation for this work was to explore the degree to which stellar systems can be used as a probe of dark matter substructures. As an example, we found that the stellar halo at $\sim 20$~kpc is expected to be sensitive to cold dark matter substructure at $k\lesssim 0.3$~kpc$^{-1}$. Perturbations are mostly driven by subhaloes larger than $10^8$~M$_\odot$, however, which may already contain visible galaxies. We also found that local disk stars may be sensitive to substructure at $k\sim 4$~kpc$^{-1}$, with perturbations driven by subhaloes as light as $\sim 10^7$~M$_\odot$. However, this conclusion depends on the extent to which subhaloes survive encounters with the Galactic disk.

These tests of the sensitivity of the stellar disk and halo relied on highly approximate descriptions of dark matter substructure. More accurate predictions of the nonlinear matter power spectrum inside a Milky Way-like halo are a subject of ongoing work. We have demonstrated in particular that it is important to know the ``spatiotemporal'' two-point statistics of the dark matter field, rather than only the spatial two-point statistics at a fixed time.

\section*{Acknowledgements}

The author thanks Jacob Nibauer for helpful comments on the manuscript.

\section*{Data Availability}
 
No new data were generated or analysed in support of this
research.



\bibliographystyle{mnras}
\bibliography{main}

\begin{thebibliography}{}
\makeatletter
\relax
\def\mn@urlcharsother{\let\do\@makeother \do\$\do\&\do\#\do\^\do\_\do\%\do\~}
\def\mn@doi{\begingroup\mn@urlcharsother \@ifnextchar [ {\mn@doi@}
  {\mn@doi@[]}}
\def\mn@doi@[#1]#2{\def\@tempa{#1}\ifx\@tempa\@empty \href
  {http://dx.doi.org/#2} {doi:#2}\else \href {http://dx.doi.org/#2} {#1}\fi
  \endgroup}
\def\mn@eprint#1#2{\mn@eprint@#1:#2::\@nil}
\def\mn@eprint@arXiv#1{\href {http://arxiv.org/abs/#1} {{\tt arXiv:#1}}}
\def\mn@eprint@dblp#1{\href {http://dblp.uni-trier.de/rec/bibtex/#1.xml}
  {dblp:#1}}
\def\mn@eprint@#1:#2:#3:#4\@nil{\def\@tempa {#1}\def\@tempb {#2}\def\@tempc
  {#3}\ifx \@tempc \@empty \let \@tempc \@tempb \let \@tempb \@tempa \fi \ifx
  \@tempb \@empty \def\@tempb {arXiv}\fi \@ifundefined
  {mn@eprint@\@tempb}{\@tempb:\@tempc}{\expandafter \expandafter \csname
  mn@eprint@\@tempb\endcsname \expandafter{\@tempc}}}

\bibitem[\protect\citeauthoryear{{Amorisco}, {G{\'o}mez}, {Vegetti}  \&
  {White}}{{Amorisco} et~al.}{2016}]{2016MNRAS.463L..17A}
{Amorisco} N.~C.,  {G{\'o}mez} F.~A.,  {Vegetti} S.,   {White} S. D.~M.,  2016,
  \mn@doi [\mnras] {10.1093/mnrasl/slw148}, \href
  {https://ui.adsabs.harvard.edu/abs/2016MNRAS.463L..17A} {463, L17}

\bibitem[\protect\citeauthoryear{{Banik} \& {Bovy}}{{Banik} \&
  {Bovy}}{2019}]{2019MNRAS.484.2009B}
{Banik} N.,  {Bovy} J.,  2019, \mn@doi [\mnras] {10.1093/mnras/stz142}, \href
  {https://ui.adsabs.harvard.edu/abs/2019MNRAS.484.2009B} {484, 2009}

\bibitem[\protect\citeauthoryear{{Banik}, {Bertone}, {Bovy}  \&
  {Bozorgnia}}{{Banik} et~al.}{2018}]{2018JCAP...07..061B}
{Banik} N.,  {Bertone} G.,  {Bovy} J.,   {Bozorgnia} N.,  2018, \mn@doi [\jcap]
  {10.1088/1475-7516/2018/07/061}, \href
  {https://ui.adsabs.harvard.edu/abs/2018JCAP...07..061B} {2018, 061}

\bibitem[\protect\citeauthoryear{{Banik}, {Bovy}, {Bertone}, {Erkal}  \& {de
  Boer}}{{Banik} et~al.}{2021a}]{2021MNRAS.502.2364B}
{Banik} N.,  {Bovy} J.,  {Bertone} G.,  {Erkal} D.,   {de Boer} T.~J.~L.,
  2021a, \mn@doi [\mnras] {10.1093/mnras/stab210}, \href
  {https://ui.adsabs.harvard.edu/abs/2021MNRAS.502.2364B} {502, 2364}

\bibitem[\protect\citeauthoryear{{Banik}, {Bovy}, {Bertone}, {Erkal}  \& {de
  Boer}}{{Banik} et~al.}{2021b}]{2021JCAP...10..043B}
{Banik} N.,  {Bovy} J.,  {Bertone} G.,  {Erkal} D.,   {de Boer} T.~J.~L.,
  2021b, \mn@doi [\jcap] {10.1088/1475-7516/2021/10/043}, \href
  {https://ui.adsabs.harvard.edu/abs/2021JCAP...10..043B} {2021, 043}

\bibitem[\protect\citeauthoryear{{Bazarov}, {Benito}, {H{\"u}tsi}, {Kipper},
  {Pata}  \& {P{\~o}der}}{{Bazarov} et~al.}{2022}]{2022A&C....4100667B}
{Bazarov} A.,  {Benito} M.,  {H{\"u}tsi} G.,  {Kipper} R.,  {Pata} J.,
  {P{\~o}der} S.,  2022, \mn@doi [Astronomy and Computing]
  {10.1016/j.ascom.2022.100667}, \href
  {https://ui.adsabs.harvard.edu/abs/2022A&C....4100667B} {41, 100667}

\bibitem[\protect\citeauthoryear{{Bland-Hawthorn} \&
  {Gerhard}}{{Bland-Hawthorn} \& {Gerhard}}{2016}]{2016ARA&A..54..529B}
{Bland-Hawthorn} J.,  {Gerhard} O.,  2016, \mn@doi [\araa]
  {10.1146/annurev-astro-081915-023441}, \href
  {https://ui.adsabs.harvard.edu/abs/2016ARA&A..54..529B} {54, 529}

\bibitem[\protect\citeauthoryear{{\VAN{Boer}{De}{de} Boer}, {Erkal}  \&
  {Gieles}}{{\VAN{Boer}{De}{de} Boer} et~al.}{2020}]{2020MNRAS.494.5315D}
{\VAN{Boer}{De}{de} Boer} T.~J.~L.,  {Erkal} D.,   {Gieles} M.,  2020, \mn@doi
  [\mnras] {10.1093/mnras/staa917}, \href
  {https://ui.adsabs.harvard.edu/abs/2020MNRAS.494.5315D} {494, 5315}

\bibitem[\protect\citeauthoryear{{Bonaca}, {Hogg}, {Price-Whelan}  \&
  {Conroy}}{{Bonaca} et~al.}{2019}]{2019ApJ...880...38B}
{Bonaca} A.,  {Hogg} D.~W.,  {Price-Whelan} A.~M.,   {Conroy} C.,  2019,
  \mn@doi [\apj] {10.3847/1538-4357/ab2873}, \href
  {https://ui.adsabs.harvard.edu/abs/2019ApJ...880...38B} {880, 38}

\bibitem[\protect\citeauthoryear{{Bovy}, {Erkal}  \& {Sanders}}{{Bovy}
  et~al.}{2017}]{2017MNRAS.466..628B}
{Bovy} J.,  {Erkal} D.,   {Sanders} J.~L.,  2017, \mn@doi [\mnras]
  {10.1093/mnras/stw3067}, \href
  {https://ui.adsabs.harvard.edu/abs/2017MNRAS.466..628B} {466, 628}

\bibitem[\protect\citeauthoryear{{Capuzzo Dolcetta}, {Di Matteo}  \&
  {Miocchi}}{{Capuzzo Dolcetta} et~al.}{2005}]{2005AJ....129.1906C}
{Capuzzo Dolcetta} R.,  {Di Matteo} P.,   {Miocchi} P.,  2005, \mn@doi [\aj]
  {10.1086/426006}, \href
  {https://ui.adsabs.harvard.edu/abs/2005AJ....129.1906C} {129, 1906}

\bibitem[\protect\citeauthoryear{{Carlberg}}{{Carlberg}}{2009}]{2009ApJ...705L.223C}
{Carlberg} R.~G.,  2009, \mn@doi [\apjl] {10.1088/0004-637X/705/2/L223}, \href
  {https://ui.adsabs.harvard.edu/abs/2009ApJ...705L.223C} {705, L223}

\bibitem[\protect\citeauthoryear{{Carlberg}}{{Carlberg}}{2012}]{2012ApJ...748...20C}
{Carlberg} R.~G.,  2012, \mn@doi [\apj] {10.1088/0004-637X/748/1/20}, \href
  {https://ui.adsabs.harvard.edu/abs/2012ApJ...748...20C} {748, 20}

\bibitem[\protect\citeauthoryear{{Carlberg}}{{Carlberg}}{2016}]{2016ApJ...820...45C}
{Carlberg} R.~G.,  2016, \mn@doi [\apj] {10.3847/0004-637X/820/1/45}, \href
  {https://ui.adsabs.harvard.edu/abs/2016ApJ...820...45C} {820, 45}

\bibitem[\protect\citeauthoryear{{Carlberg} \& {Agler}}{{Carlberg} \&
  {Agler}}{2023}]{2023arXiv230108991C}
{Carlberg} R.~G.,  {Agler} H.,  2023, \mn@doi [The Astrophysical Journal]
  {10.3847/1538-4357/ace4be}, \href
  {https://ui.adsabs.harvard.edu/abs/2023arXiv230108991C} {953, 99}

\bibitem[\protect\citeauthoryear{{Chequers}, {Widrow}  \& {Darling}}{{Chequers}
  et~al.}{2018}]{2018MNRAS.480.4244C}
{Chequers} M.~H.,  {Widrow} L.~M.,   {Darling} K.,  2018, \mn@doi [\mnras]
  {10.1093/mnras/sty2114}, \href
  {https://ui.adsabs.harvard.edu/abs/2018MNRAS.480.4244C} {480, 4244}

\bibitem[\protect\citeauthoryear{{D'Onghia}, {Springel}, {Hernquist}  \&
  {Keres}}{{D'Onghia} et~al.}{2010}]{2010ApJ...709.1138D}
{D'Onghia} E.,  {Springel} V.,  {Hernquist} L.,   {Keres} D.,  2010, \mn@doi
  [\apj] {10.1088/0004-637X/709/2/1138}, \href
  {https://ui.adsabs.harvard.edu/abs/2010ApJ...709.1138D} {709, 1138}

\bibitem[\protect\citeauthoryear{{Davies}, {Vasiliev}, {Belokurov}, {Evans}  \&
  {Dillamore}}{{Davies} et~al.}{2023}]{2023MNRAS.519..530D}
{Davies} E.~Y.,  {Vasiliev} E.,  {Belokurov} V.,  {Evans} N.~W.,   {Dillamore}
  A.~M.,  2023, \mn@doi [\mnras] {10.1093/mnras/stac3581}, \href
  {https://ui.adsabs.harvard.edu/abs/2023MNRAS.519..530D} {519, 530}

\bibitem[\protect\citeauthoryear{{Delos} \& {Schmidt}}{{Delos} \&
  {Schmidt}}{2022}]{2022MNRAS.513.3682D}
{Delos} M.~S.,  {Schmidt} F.,  2022, \mn@doi [\mnras] {10.1093/mnras/stac1022},
  \href {https://ui.adsabs.harvard.edu/abs/2022MNRAS.513.3682D} {513, 3682}

\bibitem[\protect\citeauthoryear{{Diemand}, {Kuhlen}, {Madau}, {Zemp}, {Moore},
  {Potter}  \& {Stadel}}{{Diemand} et~al.}{2008}]{2008Natur.454..735D}
{Diemand} J.,  {Kuhlen} M.,  {Madau} P.,  {Zemp} M.,  {Moore} B.,  {Potter} D.,
    {Stadel} J.,  2008, \mn@doi [\nat] {10.1038/nature07153}, \href
  {https://ui.adsabs.harvard.edu/abs/2008Natur.454..735D} {454, 735}

\bibitem[\protect\citeauthoryear{{Doke} \& {Hattori}}{{Doke} \&
  {Hattori}}{2022}]{2022ApJ...941..129D}
{Doke} Y.,  {Hattori} K.,  2022, \mn@doi [\apj] {10.3847/1538-4357/aca090},
  \href {https://ui.adsabs.harvard.edu/abs/2022ApJ...941..129D} {941, 129}

\bibitem[\protect\citeauthoryear{{Erkal} \& {Belokurov}}{{Erkal} \&
  {Belokurov}}{2015}]{2015MNRAS.454.3542E}
{Erkal} D.,  {Belokurov} V.,  2015, \mn@doi [\mnras] {10.1093/mnras/stv2122},
  \href {https://ui.adsabs.harvard.edu/abs/2015MNRAS.454.3542E} {454, 3542}

\bibitem[\protect\citeauthoryear{{Erkal}, {Belokurov}, {Bovy}  \&
  {Sanders}}{{Erkal} et~al.}{2016}]{2016MNRAS.463..102E}
{Erkal} D.,  {Belokurov} V.,  {Bovy} J.,   {Sanders} J.~L.,  2016, \mn@doi
  [\mnras] {10.1093/mnras/stw1957}, \href
  {https://ui.adsabs.harvard.edu/abs/2016MNRAS.463..102E} {463, 102}

\bibitem[\protect\citeauthoryear{{Ferguson} et~al.,}{{Ferguson}
  et~al.}{2022}]{2022AJ....163...18F}
{Ferguson} P.~S.,  et~al., 2022, \mn@doi [\aj] {10.3847/1538-3881/ac3492},
  \href {https://ui.adsabs.harvard.edu/abs/2022AJ....163...18F} {163, 18}

\bibitem[\protect\citeauthoryear{{Gonz{\'a}lez-Morales}, {Valenzuela}  \&
  {Aguilar}}{{Gonz{\'a}lez-Morales} et~al.}{2013}]{2013JCAP...03..001G}
{Gonz{\'a}lez-Morales} A.~X.,  {Valenzuela} O.,   {Aguilar} L.~A.,  2013,
  \mn@doi [\jcap] {10.1088/1475-7516/2013/03/001}, \href
  {https://ui.adsabs.harvard.edu/abs/2013JCAP...03..001G} {2013, 001}

\bibitem[\protect\citeauthoryear{{Helmi}}{{Helmi}}{2008}]{2008A&ARv..15..145H}
{Helmi} A.,  2008, \mn@doi [\aapr] {10.1007/s00159-008-0009-6}, \href
  {https://ui.adsabs.harvard.edu/abs/2008A&ARv..15..145H} {15, 145}

\bibitem[\protect\citeauthoryear{{Hernquist}}{{Hernquist}}{1990}]{1990ApJ...356..359H}
{Hernquist} L.,  1990, \mn@doi [\apj] {10.1086/168845}, \href
  {https://ui.adsabs.harvard.edu/abs/1990ApJ...356..359H} {356, 359}

\bibitem[\protect\citeauthoryear{{Ibata}, {Lewis}, {Irwin}  \& {Quinn}}{{Ibata}
  et~al.}{2002}]{2002MNRAS.332..915I}
{Ibata} R.~A.,  {Lewis} G.~F.,  {Irwin} M.~J.,   {Quinn} T.,  2002, \mn@doi
  [\mnras] {10.1046/j.1365-8711.2002.05358.x}, \href
  {https://ui.adsabs.harvard.edu/abs/2002MNRAS.332..915I} {332, 915}

\bibitem[\protect\citeauthoryear{{Ibata}, {Thomas}, {Famaey}, {Malhan},
  {Martin}  \& {Monari}}{{Ibata} et~al.}{2020}]{2020ApJ...891..161I}
{Ibata} R.,  {Thomas} G.,  {Famaey} B.,  {Malhan} K.,  {Martin} N.,   {Monari}
  G.,  2020, \mn@doi [\apj] {10.3847/1538-4357/ab7303}, \href
  {https://ui.adsabs.harvard.edu/abs/2020ApJ...891..161I} {891, 161}

\bibitem[\protect\citeauthoryear{{Johnston}, {Spergel}  \& {Haydn}}{{Johnston}
  et~al.}{2002}]{2002ApJ...570..656J}
{Johnston} K.~V.,  {Spergel} D.~N.,   {Haydn} C.,  2002, \mn@doi [\apj]
  {10.1086/339791}, \href
  {https://ui.adsabs.harvard.edu/abs/2002ApJ...570..656J} {570, 656}

\bibitem[\protect\citeauthoryear{{Koppelman} \& {Helmi}}{{Koppelman} \&
  {Helmi}}{2021}]{2021A&A...649A..55K}
{Koppelman} H.~H.,  {Helmi} A.,  2021, \mn@doi [\aap]
  {10.1051/0004-6361/202039968}, \href
  {https://ui.adsabs.harvard.edu/abs/2021A&A...649A..55K} {649, A55}

\bibitem[\protect\citeauthoryear{{K{\"u}pper}, {MacLeod}  \&
  {Heggie}}{{K{\"u}pper} et~al.}{2008}]{2008MNRAS.387.1248K}
{K{\"u}pper} A. H.~W.,  {MacLeod} A.,   {Heggie} D.~C.,  2008, \mn@doi [\mnras]
  {10.1111/j.1365-2966.2008.13323.x}, \href
  {https://ui.adsabs.harvard.edu/abs/2008MNRAS.387.1248K} {387, 1248}

\bibitem[\protect\citeauthoryear{{K{\"u}pper}, {Kroupa}, {Baumgardt}  \&
  {Heggie}}{{K{\"u}pper} et~al.}{2010}]{2010MNRAS.401..105K}
{K{\"u}pper} A. H.~W.,  {Kroupa} P.,  {Baumgardt} H.,   {Heggie} D.~C.,  2010,
  \mn@doi [\mnras] {10.1111/j.1365-2966.2009.15690.x}, \href
  {https://ui.adsabs.harvard.edu/abs/2010MNRAS.401..105K} {401, 105}

\bibitem[\protect\citeauthoryear{{K{\"u}pper}, {Lane}  \&
  {Heggie}}{{K{\"u}pper} et~al.}{2012}]{2012MNRAS.420.2700K}
{K{\"u}pper} A. H.~W.,  {Lane} R.~R.,   {Heggie} D.~C.,  2012, \mn@doi [\mnras]
  {10.1111/j.1365-2966.2011.20242.x}, \href
  {https://ui.adsabs.harvard.edu/abs/2012MNRAS.420.2700K} {420, 2700}

\bibitem[\protect\citeauthoryear{{Li} et~al.,}{{Li}
  et~al.}{2021}]{2021ApJ...911..149L}
{Li} T.~S.,  et~al., 2021, \mn@doi [\apj] {10.3847/1538-4357/abeb18}, \href
  {https://ui.adsabs.harvard.edu/abs/2021ApJ...911..149L} {911, 149}

\bibitem[\protect\citeauthoryear{{Malhan}, {Valluri}  \& {Freese}}{{Malhan}
  et~al.}{2021}]{2021MNRAS.501..179M}
{Malhan} K.,  {Valluri} M.,   {Freese} K.,  2021, \mn@doi [\mnras]
  {10.1093/mnras/staa3597}, \href
  {https://ui.adsabs.harvard.edu/abs/2021MNRAS.501..179M} {501, 179}

\bibitem[\protect\citeauthoryear{{Montanari} \&
  {Garc{\'\i}a-Bellido}}{{Montanari} \&
  {Garc{\'\i}a-Bellido}}{2022}]{2022PDU....3500978M}
{Montanari} F.,  {Garc{\'\i}a-Bellido} J.,  2022, \mn@doi [Physics of the Dark
  Universe] {10.1016/j.dark.2022.100978}, \href
  {https://ui.adsabs.harvard.edu/abs/2022PDU....3500978M} {35, 100978}

\bibitem[\protect\citeauthoryear{{Ngan} \& {Carlberg}}{{Ngan} \&
  {Carlberg}}{2014}]{2014ApJ...788..181N}
{Ngan} W.~H.~W.,  {Carlberg} R.~G.,  2014, \mn@doi [\apj]
  {10.1088/0004-637X/788/2/181}, \href
  {https://ui.adsabs.harvard.edu/abs/2014ApJ...788..181N} {788, 181}

\bibitem[\protect\citeauthoryear{{Pe{\~n}arrubia}}{{Pe{\~n}arrubia}}{2019}]{2019MNRAS.484.5409P}
{Pe{\~n}arrubia} J.,  2019, \mn@doi [\mnras] {10.1093/mnras/stz338}, \href
  {https://ui.adsabs.harvard.edu/abs/2019MNRAS.484.5409P} {484, 5409}

\bibitem[\protect\citeauthoryear{{Pearson}, {Price-Whelan}  \&
  {Johnston}}{{Pearson} et~al.}{2017}]{2017NatAs...1..633P}
{Pearson} S.,  {Price-Whelan} A.~M.,   {Johnston} K.~V.,  2017, \mn@doi [Nature
  Astronomy] {10.1038/s41550-017-0220-3}, \href
  {https://ui.adsabs.harvard.edu/abs/2017NatAs...1..633P} {1, 633}

\bibitem[\protect\citeauthoryear{{Penarrubia}, {Koposov}, {Walker}, {Gilmore},
  {Wyn Evans}  \& {Mackay}}{{Penarrubia} et~al.}{2010}]{2010arXiv1005.5388P}
{Penarrubia} J.,  {Koposov} S.~E.,  {Walker} M.~G.,  {Gilmore} G.,  {Wyn Evans}
  N.,   {Mackay} C.~D.,  2010, \mn@doi [arXiv e-prints]
  {10.48550/arXiv.1005.5388}, \href
  {https://ui.adsabs.harvard.edu/abs/2010arXiv1005.5388P} {p. arXiv:1005.5388}

\bibitem[\protect\citeauthoryear{{Qian}, {Arshad}  \& {Bovy}}{{Qian}
  et~al.}{2022}]{2022MNRAS.511.2339Q}
{Qian} Y.,  {Arshad} Y.,   {Bovy} J.,  2022, \mn@doi [\mnras]
  {10.1093/mnras/stac238}, \href
  {https://ui.adsabs.harvard.edu/abs/2022MNRAS.511.2339Q} {511, 2339}

\bibitem[\protect\citeauthoryear{{Ramirez} \& {Buckley}}{{Ramirez} \&
  {Buckley}}{2023}]{2023MNRAS.525.5813R}
{Ramirez} E.~D.,  {Buckley} M.~R.,  2023, \mn@doi [\mnras]
  {10.1093/mnras/stad2583}, \href
  {https://ui.adsabs.harvard.edu/abs/2023MNRAS.525.5813R} {525, 5813}

\bibitem[\protect\citeauthoryear{{Sanders}, {Bovy}  \& {Erkal}}{{Sanders}
  et~al.}{2016}]{2016MNRAS.457.3817S}
{Sanders} J.~L.,  {Bovy} J.,   {Erkal} D.,  2016, \mn@doi [\mnras]
  {10.1093/mnras/stw232}, \href
  {https://ui.adsabs.harvard.edu/abs/2016MNRAS.457.3817S} {457, 3817}

\bibitem[\protect\citeauthoryear{{Scherrer} \& {Bertschinger}}{{Scherrer} \&
  {Bertschinger}}{1991}]{1991ApJ...381..349S}
{Scherrer} R.~J.,  {Bertschinger} E.,  1991, \mn@doi [\apj] {10.1086/170658},
  \href {https://ui.adsabs.harvard.edu/abs/1991ApJ...381..349S} {381, 349}

\bibitem[\protect\citeauthoryear{{Siegal-Gaskins} \&
  {Valluri}}{{Siegal-Gaskins} \& {Valluri}}{2008}]{2008ApJ...681...40S}
{Siegal-Gaskins} J.~M.,  {Valluri} M.,  2008, \mn@doi [\apj] {10.1086/587450},
  \href {https://ui.adsabs.harvard.edu/abs/2008ApJ...681...40S} {681, 40}

\bibitem[\protect\citeauthoryear{{Starkenburg} \& {Helmi}}{{Starkenburg} \&
  {Helmi}}{2015}]{2015A&A...575A..59S}
{Starkenburg} T.~K.,  {Helmi} A.,  2015, \mn@doi [\aap]
  {10.1051/0004-6361/201425082}, \href
  {https://ui.adsabs.harvard.edu/abs/2015A&A...575A..59S} {575, A59}

\bibitem[\protect\citeauthoryear{{Tavangar} et~al.,}{{Tavangar}
  et~al.}{2022}]{2022ApJ...925..118T}
{Tavangar} K.,  et~al., 2022, \mn@doi [\apj] {10.3847/1538-4357/ac399b}, \href
  {https://ui.adsabs.harvard.edu/abs/2022ApJ...925..118T} {925, 118}

\bibitem[\protect\citeauthoryear{{Tremaine}, {Frankel}  \& {Bovy}}{{Tremaine}
  et~al.}{2023}]{2023MNRAS.521..114T}
{Tremaine} S.,  {Frankel} N.,   {Bovy} J.,  2023, \mn@doi [\mnras]
  {10.1093/mnras/stad577}, \href
  {https://ui.adsabs.harvard.edu/abs/2023MNRAS.521..114T} {521, 114}

\bibitem[\protect\citeauthoryear{{Weatherford}, {Rasio}, {Chatterjee},
  {Fragione}, {K{\i}ro{\u{g}}lu}  \& {Kremer}}{{Weatherford}
  et~al.}{2023}]{2023arXiv231001485W}
{Weatherford} N.~C.,  {Rasio} F.~A.,  {Chatterjee} S.,  {Fragione} G.,
  {K{\i}ro{\u{g}}lu} F.,   {Kremer} K.,  2023, \mn@doi [arXiv e-prints]
  {10.48550/arXiv.2310.01485}, \href
  {https://ui.adsabs.harvard.edu/abs/2023arXiv231001485W} {p. arXiv:2310.01485}

\bibitem[\protect\citeauthoryear{{Webb}, {Bovy}, {Carlberg}  \&
  {Gieles}}{{Webb} et~al.}{2019}]{2019MNRAS.488.5748W}
{Webb} J.~J.,  {Bovy} J.,  {Carlberg} R.~G.,   {Gieles} M.,  2019, \mn@doi
  [\mnras] {10.1093/mnras/stz2118}, \href
  {https://ui.adsabs.harvard.edu/abs/2019MNRAS.488.5748W} {488, 5748}

\bibitem[\protect\citeauthoryear{{Widrow}}{{Widrow}}{2023}]{2023MNRAS.522..477W}
{Widrow} L.~M.,  2023, \mn@doi [\mnras] {10.1093/mnras/stad973}, \href
  {https://ui.adsabs.harvard.edu/abs/2023MNRAS.522..477W} {522, 477}

\bibitem[\protect\citeauthoryear{{Yoon}, {Johnston}  \& {Hogg}}{{Yoon}
  et~al.}{2011}]{2011ApJ...731...58Y}
{Yoon} J.~H.,  {Johnston} K.~V.,   {Hogg} D.~W.,  2011, \mn@doi [\apj]
  {10.1088/0004-637X/731/1/58}, \href
  {https://ui.adsabs.harvard.edu/abs/2011ApJ...731...58Y} {731, 58}

\makeatother
\end{thebibliography}



\appendix


\bsp	
\label{lastpage}
\end{document}